\documentclass[preprint,amsmath,amssymb,aps,showkeys,showpacs]{revtex4}
\usepackage[english]{babel}
\usepackage{graphicx}
\usepackage{graphics}
\usepackage{amsmath}
\usepackage{dcolumn}
\usepackage{amssymb}
\usepackage{bm}

\begin{document}
\title{Torsion effects on a relativistic position-dependent mass system}
\author{R. L. L. Vit\'oria}
\affiliation{Departamento de F\'isica, Universidade Federal da Para\'iba, Caixa Postal 5008, 58051-900, Jo\~ao Pessoa-PB, Brazil.} 

\author{K. Bakke}
\email{kbakke@fisica.ufpb.br} 
\affiliation{Departamento de F\'isica, Universidade Federal da Para\'iba, Caixa Postal 5008, 58051-900, Jo\~ao Pessoa-PB, Brazil.}

\begin{abstract}
We analyse a relativistic scalar particle with a position-dependent mass in a spacetime with a space-like dislocation by showing that relativistic bound states solutions can be achieved. Further, we consider the presence of the Coulomb potential and analyse the relativistic position-dependent mass system subject to the Coulomb potential in the spacetime with a space-like dislocation. We also show that a new set of relativistic bound states solutions can be obtained, where there also exists the influence of torsion of the relativistic energy levels. Finally, we investigate an analogue of the Aharonov-Bohm effect for bound states in this position-dependent mass in a spacetime with a space-like dislocation.
\end{abstract}

\keywords{Topological defect, torsion, dislocation, position-dependent mass, relativistic bound states solutions, Aharonov-Bohm effect}
\pacs{03.65.Pm, 03.65.Ge, 11.27.+d}

\maketitle

\section{Introduction}

Position-dependent mass systems have have attracted a great deal of attention in condensed matter physics due to the developments achieved in recent decades in studies of semiconductors \cite{pdm6,pdm7,pdm8}, quantum liquids \cite{pdm9} and quantum dots \cite{pdm4}. At present days, several discussions can be found in the literature with the purpose of showing different ways of dealing with position dependent-mass systems \cite{pdm,pdm1,pdm1a,pdm1b,pdm1c,pdm2,pdm3,pdm5,pdm10,pdm11,pdm12,pdm14,pdm17,pdm19,ex1,ex2}. Another context has been discussed in Ref. \cite{greiner} that has a particular interest in quantum field theory \cite{quark,pdm22}. In that case, a position-dependent mass system can be built by introducing a scalar potential as a modification in the mass term of the Klein-Gordon equation. In this way, the quark-antiquark interaction has been dealt with in Ref. \cite{quark} as a position-dependent mass system. It is worth mentioning other interests in position-dependent mass systems, such as, in $\mathcal{PT}$-symmetric quantum mechanics \cite{pdm13,pdm21,pdm23}, supersymmetric quantum mechanics \cite{pdm25,pdm26,pdm27} and relativistic quantum mechanics \cite{pdm15,pdm16,pdm18,pdm20,pdm24,bf40,vb}.

In this work, we investigate relativistic quantum effects on a scalar particle that possesses a position-dependent mass in a spacetime with a space-like dislocation. It is worth emphasizing that a spacetime with a space-like dislocation corresponds to a topological defect spacetime associated with the presence of torsion in the spacetime \cite{put}. In quantum mechanics, the background defined by a spacetime with a space-like dislocation has been explored in studies of the Aharonov-Bohm effect for bound states \cite{valdir}, quantum scattering \cite{valdir2,fur3}, harmonic oscillator \cite{fur2}, noninertial effects \cite{b}, Kaluza-Klein theory \cite{fur4} and the Dirac oscillator \cite{bf2}.

The structure of this paper is as follows: in section II, we introduce a relativistic position-dependent mass system in a topological defect spacetime, then, we show that we can solve the Klein-Gordon equation analytically and discuss the influence of torsion on the relativistic energy levels; in section III, we analyse the influence of the Coulomb interaction and torsion on the relativistic position-dependent mass system by obtaining the relativistic bound states solutions. In particular, we calculate the allowed energies for the lowest energy state of the system; in section IV, we investigate the influence of torsion and the Aharonov-Bohm quantum flux \cite{ab} on the relativistic position-dependent mass system. We show that relativistic bound states solutions can be achieved. In particular, we obtain the relativistic energy levels of the lowest energy state of the system and show the dependence of these energy levels on the the Aharonov-Bohm quantum flux \cite{ab} and the parameter associated with torsion. From this dependence of the relativistic energy levels on the geometric quantum phase, we calculate the persistent current associated with the lowest energy state of the system. In section V, we present ours conclusions.

\section{Relativistic position-dependent mass particle in a spacetime with a space-like dislocation}

In this section, we analyse the behaviour of a relativistic scalar particle with a position-dependent mass in a spacetime with a space-like dislocation. This position-dependent mass system is built by introducing a scalar potential as a modification in the mass term of the Klein-Gordon equation as: $m\rightarrow m+S\left(\vec{r}\right)$, where $S\left(\vec{r}\right)$ is the scalar potential \cite{greiner}. In the present work, we assume that the relativistic scalar particle possesses a position-dependent mass given by
\begin{eqnarray}
m\left(\rho\right)=m+\nu\,\rho,
\label{1.1}
\end{eqnarray}
where $\nu$ is a constant parameter that characterizes the scalar potential $S\left(\vec{r}\right)=\nu\,\rho$, where $\rho=\sqrt{x^{2}+y^{2}}$ is the radial coordinate. Therefore, the relativistic quantum description of this position-dependent mass system is given by the Klein-Gordon equation:
\begin{eqnarray}
\hat{p}^{\mu}\,\hat{p}_{\mu}\phi-\left(m+\nu\rho\right)^{2}\,\phi=0.
\label{1.2}
\end{eqnarray}
Now, let us consider a spacetime with a space-like dislocation, where the line element is given in the form \cite{put,vb}:
\begin{eqnarray}
ds^{2}=-dt^{2}+d\rho^{2}+\rho^{2}\,d\varphi^{2}+\left(dz+\chi\,d\varphi\right)^{2}.
\label{1.3}
\end{eqnarray}
In Eq. (\ref{1.3}), the parameter $\chi$ is a constant parameter related to torsion, where the torsion corresponds to a singularity at the origin \cite{fur2,fur3,put,vb}. In contrast to the elastic theory in condensed matter physics \cite{kleinert,kat}, the parameter $\chi$ is related to the Burgers vector $\vec{b}$ via $\chi=b/2\pi$. Thereby, the Klein-Gordon equation (\ref{1.2}) in the spacetime with a space-like dislocation (\ref{1.2}), becomes
\begin{eqnarray}
\left(m+\nu\rho\right)^{2}\phi=-\frac{\partial^{2}\phi}{\partial t^{2}}+\frac{\partial^{2}\phi}{\partial\rho^{2}}+\frac{1}{\rho}\frac{\partial\phi}{\partial\rho}+\frac{1}{\rho^{2}}\left(\frac{\partial}{\partial\varphi}-\chi\frac{\partial}{\partial z}\right)^{2}\phi+\frac{\partial^{2}\phi}{\partial z^{2}}.
\label{1.4}
\end{eqnarray}
A solution to Eq. (\ref{1.4}) can be written through the ansatz:
\begin{eqnarray}
\phi\left(t,\,\rho,\,\varphi,\,z\right)=e^{-i\mathcal{E}t}\,e^{il\varphi}\,e^{ikz}\,R\left(\rho\right),
\label{1.5}
\end{eqnarray}
where $l=0,\pm1,\pm2,\ldots$ and $R\left(\rho\right)$ is a function of the radial coordinate. Then, we have
\begin{eqnarray}
R''+\frac{1}{\rho}\,R'-\frac{1}{\rho^{2}}\left(l-\chi\,k\right)^{2}R-2m\,\nu\,\rho\,R-\nu^{2}\,\rho^{2}R+\left(\mathcal{E}^{2}-m^{2}-k^{2}\right)R=0.
\label{1.6}
\end{eqnarray}

Let us perform a change of variables given by $\xi=\sqrt{\nu}\,\rho$, hence, the radial equation (\ref{1.6}) becomes
\begin{eqnarray}
R''+\frac{1}{\xi}\,R'-\frac{\gamma^{2}}{\xi^{2}}R-\xi^{2}\,R-\alpha\,\xi\,R+\beta\,R=0,
\label{1.7}
\end{eqnarray}
where we have defined $\gamma=l-\chi\,k$, $\alpha=\frac{2m}{\sqrt{\nu}}$ and $\beta=\frac{1}{\nu}\left(\mathcal{E}^{2}-m^{2}-k^{2}\right)$. By analysing the asymptotic behaviour at $\xi\rightarrow0$ and $\xi\rightarrow\infty$, then, we can write the function $R\left(\xi\right)$ in terms of an unknown function $G\left(\xi\right)$ in the form \cite{eug,mhv,vercin,heun}:
\begin{eqnarray}
R\left(\xi\right)=e^{-\frac{\xi^{2}}{2}}\,e^{-\frac{\alpha\,\xi}{2}}\,\xi^{\left|\gamma\right|}\,G\left(\xi\right),
\label{1.10}
\end{eqnarray}
and thus, by substituting Eq. (\ref{1.10}) into Eq. (\ref{1.7}), we obtain that the function $G\left(\xi\right)$ is a solution to the following second order differential equation:
\begin{eqnarray}
G''+\left[\frac{2\left|\gamma\right|+1}{\xi}-\alpha-2\xi\right]G'+\left[\beta+\frac{\alpha^{2}}{4}-2-2\left|\gamma\right|-\frac{\alpha\left[2\left|\gamma\right|+1\right]}{2\xi}\right]G=0.
 \label{1.11}
\end{eqnarray}
This second order differential equation is called as the biconfluent Heun equation \cite{heun}, and the function $G\left(\xi\right)=H_{B}\left(2\left|\gamma\right|,\,\alpha,\,\beta+\frac{\alpha^{2}}{4},\,0,\,\xi\right)$ is the biconfluent Heun function.

Let us search for polynomial solutions to Eq. (\ref{1.11}), then, for this purpose, we write the solution to Eq. (\ref{1.11}) as a power series expansion around the origin: $G\left(\xi\right)=\sum_{k=0}^{\infty}a_{k}\,\xi^{k}$ \cite{arf,eug}. Hence, by substituting this series into Eq. (\ref{1.11}), we obtain the recurrence relation
\begin{eqnarray}
a_{k+2}=\frac{\alpha\left(k+1\right)+\tau}{\left(k+2\right)\left(k+2+2\left|\gamma\right|\right)}\,a_{k+1}-\frac{\lambda-2k}{\left(k+2\right)\left(k+2+2\left|\gamma\right|\right)}\,a_{k},
\label{1.13}
\end{eqnarray}
and the relation
\begin{eqnarray}
a_{1}=\frac{\alpha}{2}\,a_{0},
\label{1.13a}
\end{eqnarray}
where 
\begin{eqnarray}
\lambda=\beta+\frac{\alpha^{2}}{4}-2-2\left|\gamma\right|;\,\,\,\tau=\frac{\alpha}{2}\left[2\left|\gamma\right|+1\right].
\label{1.14}
\end{eqnarray}

Note that polynomial solutions to the function $G\left(\xi\right)$ are achieved by imposing that the biconfluent Heun series becomes a polynomial of degree $n$. From the recurrence relation (\ref{1.13}), we have that the biconfluent Heun series becomes a polynomial of degree $n$ by imposing that \cite{bf40,eug}:
\begin{eqnarray}
\lambda=2n;\,\,\,\,\,a_{n+1}=0,
\label{1.15}
\end{eqnarray}
where $n=1,2,3,\ldots$. With the condition $\lambda=2n$, we obtain
\begin{eqnarray}
\mathcal{E}_{n,\,l,\,k}=\pm\sqrt{2\nu\left[n+\left|l-\chi\,k\right|+1\right]+k^{2}}.
\label{1.16}
\end{eqnarray}

On the other hand, if we wish to analyse the condition $a_{n+1}=0$, we need to obtain some coefficients of the power series expansion. Let us start with $a_{0}=1$, then, from Eqs. (\ref{1.13}) and (\ref{1.13a}) we obtain $a_{1}=\frac{\alpha}{2}$ and $a_{2}=\frac{\alpha^{2}\left(2\left|\gamma\right|+3\right)}{8\left(2+2\left|\gamma\right|\right)}-\frac{\lambda}{2\left(2+2\left|\gamma\right|\right)}$. In this way, if we take the lowest energy state $\left(n=1\right)$, we have $a_{n+1}=a_{2}=0$, and then,
\begin{eqnarray}
\nu_{1,\,l,\,k}=m^{2}\left(\left|l-\chi\,k\right|+\frac{3}{2}\right),
\label{1.17}
\end{eqnarray} 
i.e., in order to achieve polynomial solutions to the function $G\left(r\right)$, we have assumed that the parameter $\nu$ associated with the linear scalar potential in Eq. (\ref{1.1}) must be chosen with the purpose of satisfying the condition $a_{n+1}=0$, therefore, we have labeled $\nu=\nu_{n,\,l,\,k}$ in Eq. (\ref{1.17}). With the relation given in Eq. (\ref{1.17}), we have that the possible values of the parameter $\nu$ are determined by the parameter associated with torsion and the quantum numbers $\left\{n,\,l,\,k\right\}$ of the system. By substituting Eq. (\ref{1.17}) into Eq. (\ref{1.16}), the allowed energies for the lowest energy state are given by
\begin{eqnarray}
\mathcal{E}_{1,\,l,\,k}=\pm\,m\,\sqrt{\left(2\left|l-\chi\,k\right|+3\right)\left(\left|l-\chi\,k\right|+2\right)+\frac{k^{2}}{m^{2}}}.
\label{1.18}
\end{eqnarray}

Hence, for each relativistic energy level we have a different expression for $\nu$ and, thus, we can rewrite Eq. (\ref{1.16}) as
\begin{eqnarray}
\mathcal{E}_{n,\,l,\,k}=\pm\sqrt{2\nu_{n,\,l,\,k}\left[n+\left|l-\chi\,k\right|+1\right]+k^{2}},
\label{1.19}
\end{eqnarray}
which is the general expression for the relativistic energy levels for the position-dependent mass system described in Eq. (\ref{1.1}) in the spacetime with a space-like dislocation. Note that we can observe the influence of torsion in Eqs. (\ref{1.18}) and (\ref{1.19}) through the presence of the parameter $\chi$, which is the constant parameter related to torsion. It yields a shift in the angular momentum that gives rise to an effective angular momentum quantum number $l_{\mathrm{eff}}=l-\chi\,k$. As pointed out in Ref. \cite{valdir}, this shift is analogous to the Aharonov-Bohm effect \cite{ab,pesk}, where the shift in the angular momentum is given by $\bar{l}=l-e\Phi/2\pi$. By taking $\chi=0$, then, we obtain in Eq. (\ref{1.19}) the relativistic energy levels for the position-dependent mass system in the Minkowski spacetime.

\section{Effects of the Coulomb interaction on the relativistic position-dependent mass system}

On the other hand, the Coulomb potential can be introduced into the Klein-Gordon equation through a minimal coupling $\hat{p}_{\mu}\rightarrow\hat{p}_{\mu}-q\,A_{\mu}\left(x\right)$ \cite{greiner}, where $q$ is the electric charge and $A_{\mu}=\left(-A_{0},\,\vec{A}\right)$ is the electromagnetic 4-vector potential. Then, in this section, we analyse the relativistic position-dependent mass system given in Eq. (\ref{1.1}) subject to the Coulomb potential in the spacetime with a space-like dislocation (\ref{1.3}). Thereby, the corresponding Klein-Gordon equation is:
\begin{eqnarray}
\left[\hat{p}^{\mu}-q\,A^{\mu}\left(x\right)\right]\left[\hat{p}_{\mu}-q\,A_{\mu}\left(x\right)\right]\phi-\left(m+\nu\rho\right)^{2}\,\phi=0.
\label{2.1}
\end{eqnarray}
From now on, we write the term associated with the Coulomb potential as $q\,A_{0}=\frac{b}{\rho}=\pm\frac{\left|b\right|}{\rho}$, where $b$ is a constant. Then, the Klein-Gordon equation (\ref{1.2}) in the spacetime with a space-like dislocation (\ref{1.3}),becomes
\begin{eqnarray}
\left(m+\nu\rho\right)^{2}\phi=-\frac{\partial^{2}\phi}{\partial t^{2}}+i\frac{2b}{\rho}\frac{\partial\phi}{\partial t}+\frac{b^{2}}{\rho^{2}}\,\phi+\frac{\partial^{2}\phi}{\partial\rho^{2}}+\frac{1}{\rho}\frac{\partial\phi}{\partial\rho}+\frac{1}{\rho^{2}}\left(\frac{\partial}{\partial\varphi}-\chi\frac{\partial}{\partial z}\right)^{2}\phi+\frac{\partial^{2}\phi}{\partial z^{2}}.
\label{2.2}
\end{eqnarray}

By using the same ansatz given in (\ref{1.5}) and by performing the change of variables $\xi=\sqrt{\nu}\,\rho$ as in the previous section, we obtain the second order differential equation:
\begin{eqnarray}
R''+\frac{1}{\xi}\,R'-\frac{\eta^{2}}{\xi^{2}}\,R+\frac{\mu}{\xi}\,R-\alpha\,\xi\,R-\xi^{2}\,R+\beta\,R=0,
\label{2.3}
\end{eqnarray}
where $\eta^{2}=\left(l-\chi\,k\right)^{2}+b^{2}$, $\mu=\frac{2\,b\,\mathcal{E}}{\sqrt{\nu}}$, $\alpha=\frac{2m}{\sqrt{\nu}}$ and $\beta=\frac{1}{\nu}\left(\mathcal{E}^{2}-m^{2}-k^{2}\right)$. In order to obtain a radial wave function in which is well-behaved at $\xi\rightarrow0$ and $\xi\rightarrow\infty$, then, we write the solution to Eq. (\ref{2.3}) in the form:
\begin{eqnarray}
R\left(\xi\right)=e^{-\frac{\xi^{2}}{2}}\,e^{-\frac{\alpha\,\xi}{2}}\,\xi^{\left|\eta\right|}\,\bar{G}\left(\xi\right),
\label{2.4}
\end{eqnarray}
where $\bar{G}\left(\xi\right)$ is an unknown function. By substituting Eq. (\ref{2.4}) into Eq. (\ref{2.3}), we obtain
\begin{eqnarray}
\bar{G}''+\left[\frac{2\left|\eta\right|+1}{\xi}-\alpha-2\xi\right]\bar{G}'+\left[\beta+\frac{\alpha^{2}}{4}-2-2\left|\eta\right|-\frac{\alpha\left[2\left|\eta\right|+1\right]-\mu}{2\xi}\right]\bar{G}=0,
\label{2.5}
\end{eqnarray}
which is also the biconfluent Heun equation \cite{heun} and the $\bar{G}\left(\xi\right)=H_{B}\left(2\left|\eta\right|,\,\alpha,\,\beta+\frac{\alpha^{2}}{4},\,-2\mu,\,\xi\right)$ is the biconfluent Heun function. By following the steps from Eq. (\ref{1.13}) to Eq. (\ref{1.15}), we can write $\bar{G}\left(\xi\right)=\sum_{k=0}^{\infty}d_{k}\,\xi^{k}$ \cite{arf}, and then, we obtain the recurrence relation
\begin{eqnarray}
d_{k+2}=\frac{\alpha\left(k+1\right)+\bar{\tau}}{\left(k+2\right)\left(k+2+2\left|\eta\right|\right)}\,d_{k+1}-\frac{\bar{\lambda}-2k}{\left(k+2\right)\left(k+2+2\left|\eta\right|\right)}\,d_{k},
\label{2.6}
\end{eqnarray}
and we also have
\begin{eqnarray}
d_{1}=\frac{\bar{\tau}}{\left(1+2\left|\eta\right|\right)}\,d_{0},
\label{2.6a}
\end{eqnarray}
where 
\begin{eqnarray}
\bar{\lambda}=\beta+\frac{\alpha^{2}}{4}-2-2\left|\eta\right|;\,\,\,\bar{\tau}=\frac{\alpha}{2}\left[2\left|\eta\right|+1\right]-\mu.
\label{2.7}
\end{eqnarray}

By starting with $d_{0}=1$, we obtain with Eqs. (\ref{2.6}) and (\ref{2.6a}) that $d_{1}=\frac{\bar{\tau}}{\left(1+2\left|\eta\right|\right)}$ and $d_{2}=\frac{\bar{\tau}\left(\alpha+\bar{\tau}\right)}{2\left(2+2\left|\eta\right|\right)\left(1+2\left|\eta\right|\right)}-\frac{\bar{\lambda}}{2\left(2+2\left|\eta\right|\right)}$. Again, we have that polynomial solutions to the function $\bar{G}\left(\xi\right)$ are achieved by imposing that the biconfluent Heun series becomes a polynomial of degree $n$. From the recurrence relation (\ref{2.6}), we need to impose that $\bar{\lambda}=2n$ and $d_{n+1}=0$, with $n=1,2,3,\ldots$. With the condition $\bar{\lambda}=2n$, we obtain the general expression for the relativistic energy levels:
\begin{eqnarray}
\mathcal{E}_{n,\,l,\,k}=\pm\sqrt{2\nu_{n,\,l,\,k}\left[n+\left|\eta\right|+1\right]+k^{2}},
\label{2.8}
\end{eqnarray}
where we have labelled $\nu=\nu_{n,\,l,\,k}$ as in the previous section and defined $\eta^{2}=\left(l-\chi\,k\right)^{2}+b^{2}$. Further, let us analyse the condition $d_{n+1}=0$ by working with the lowest energy state $n=1$. In this case, we have that $d_{n+1}=d_{2}=0$, and then, the possible values of the parameter $\nu$ are determined by
\begin{eqnarray}
\nu_{1,\,l,\,k}=\frac{m^{2}}{2}\left(2\left|\eta\right|+3\right)-2m\,b\,\mathcal{E}_{1,\,l}\,\frac{\left(2\left|\eta\right|+2\right)}{\left(2\left|\eta\right|+1\right)}+\frac{2\,b^{2}\,\mathcal{E}_{1,\,l}^{2}}{\left(2\left|\eta\right|+1\right)},
\label{2.9}
\end{eqnarray}
where we can see that the possible values of $\nu$ are determined by the parameters associated with torsion and the Coulomb potential, and then, by the quantum numbers of the system $\left\{n,\,l,\,k\right\}$. With the result given in Eq. (\ref{2.9}), hence, the allowed energies for the lowest energy state are
\begin{eqnarray}
\mathcal{E}_{1,\,l,\,k}&=&\frac{2m\,b\left(\left|\eta\right|+2\right)\left(2\left|\eta\right|+2\right)}{\left(4b^{2}\left|\eta\right|+8b^{2}-2\left|\eta\right|-1\right)}\times\nonumber\\
[-2mm]\label{2.10}\\[-2mm]
&\times&\left[1\pm\sqrt{1-\frac{\left(4b^{2}\left|\eta\right|+8b^{2}-2\left|\eta\right|-1\right)\left[m^{2}\left(\left|\eta\right|+2\right)\left(2\left|\eta\right|+3\right)+k^{2}\right]\left(2\left|\eta\right|+1\right)}{4m^{2}b^{2}\left(\left|\eta\right|+2\right)^{2}\left(2\left|\eta\right|+2\right)^{2}}}\right].\nonumber
\end{eqnarray}

Hence, Eq. (\ref{2.8}) corresponds to the general expression for the relativistic energy levels of the position-dependent mass system given in Eq. (\ref{1.1}) under the influence of the Coulomb potential in the spacetime with a space-like dislocation. By comparing the expressions for the relativistic energy levels (\ref{2.8}) and (\ref{2.10}) with that ones obtained in the previous section, we have that the presence of the Coulomb potential modifies the spectrum of energy of the relativistic position-dependent mass system. Besides, from Eqs. (\ref{2.8}) and (\ref{2.10}), we have that torsion effects exist on the relativistic energy levels. Note that by taking $\chi=0$ in Eqs. (\ref{2.8}) and (\ref{2.10}), we obtain the general expression of the relativistic energy levels and the allowed energies of the lowest energy state in the Minkowski spacetime.

\section{Aharonov-Bohm effect for bound states }

In this section, let us consider the electromagnetic 4-vector potential to be given by the component $A_{\varphi}=\frac{\Phi_{B}}{2\pi}$, where $\Phi_{B}$ denotes the Aharonov-Bohm quantum flux \cite{ab,eug2}. It is worth mentioning other works that have investigated the Aharonov-Bohm effect \cite{ex3,ex4,ex5,ex6,ex7,ex8,ex9,ex10,ex11,ex12,ex13}. In this way, the Klein-Gordon equation (\ref{2.1}) becomes:
\begin{eqnarray}
\left(m+\nu\rho\right)^{2}\phi=-\frac{\partial^{2}\phi}{\partial t^{2}}+\frac{\partial^{2}\phi}{\partial\rho^{2}}+\frac{1}{\rho}\frac{\partial\phi}{\partial\rho}+\frac{1}{\rho^{2}}\left(\frac{\partial}{\partial\varphi}-\chi\frac{\partial}{\partial z}+i\,q\frac{\Phi_{B}}{2\pi}\right)^{2}\phi+\frac{\partial^{2}\phi}{\partial z^{2}}.
\label{3.1}
\end{eqnarray}
 
Note that, by replacing $\gamma=l-\chi\,k$ with $\varsigma=l-\chi\,k+\frac{q\,\Phi_{B}}{2\pi}$, we can follow the same steps from Eq. (\ref{1.5}) to Eq. (\ref{1.19}) in order to solve the Klein-Gordon equation (\ref{3.1}). In this way, the general expression for the relativistic energy levels for the position-dependent mass system described in Eq. (\ref{1.1}) in the spacetime with a space-like dislocation becomes
\begin{eqnarray}
\mathcal{E}_{n,\,l,\,k}=\pm\sqrt{2\nu_{n,\,l,\,k}\left[n+\left|\varsigma\right|+1\right]+k^{2}},
\label{3.2}
\end{eqnarray}
where there exists the influence of the Aharonov-Bohm geometric quantum phase $\Phi_{B}$ and the topological defect on the relativistic energy levels. This dependence of the relativistic energy levels on the geometric quantum phase corresponds to a relativistic analogue of the Aharonov-Bohm effect for bound states \cite{pesk,bf2,bf3}. Furthermore, by dealing with the lowest energy state of the system, we follow the steps from Eq. (\ref{1.13}) to Eq. (\ref{1.17}) and obtain
\begin{eqnarray}
\nu_{1,\,l,\,k}=m^{2}\left(\left|l-\chi\,k+\frac{q\,\Phi_{B}}{2\pi}\right|+\frac{3}{2}\right).
\label{3.3}
\end{eqnarray}
Hence, the possible values of the parameter $\nu$ depend on the Aharonov-Bohm geometric quantum phase \cite{ab}, the parameter associated with torsion and the quantum numbers $\left\{n,\,l,\,k\right\}$ of the system in order that polynomial solutions to the radial wave function can be achieved. By substituting Eq. (\ref{3.3}), the allowed energies for the lowest energy state are given by
\begin{eqnarray}
\mathcal{E}_{1,\,l,\,k}=\pm\,m\,\sqrt{\left(2\left|l-\chi\,k+\frac{q\,\Phi_{B}}{2\pi}\right|+3\right)\left(\left|l-\chi\,k+\frac{q\,\Phi_{B}}{2\pi}\right|+2\right)+\frac{k^{2}}{m^{2}}}.
\label{3.4}
\end{eqnarray}

It is worth noting that there exists torsion effects on both the general expression of the relativistic energy levels (\ref{3.2}) and the particular case obtained in Eq. (\ref{3.4}). However, by taking $\chi=0$ in Eqs. (\ref{3.2})-(\ref{3.4}), we obtain the relativistic analogue of the Aharonov-Bohm effect for bound states in the Minkowski spacetime. On the other hand, by taking $\Phi_{B}=0$ in Eqs. (\ref{3.2}) and (\ref{3.4}), we recover the results obtained in Eqs. (\ref{1.16})-(\ref{1.19}).

In recent years, it has been discussed the dependence of the energy levels on the geometric quantum phase in relativistic quantum systems \cite{,fur4,bf2,bf3,lbb} and an analogue of a well-known effect in condensed matter systems called as the persistent currents \cite{by,tan,fur5,per1,per2}. It can be found in the literature studies of persistent currents associated with the Berry phase \cite{per3,per4}, the Aharonov-Anandan quantum phase \cite{per5,per6} and the Aharonov-Casher geometric quantum phase \cite{per7,per8,per9,per10}. According to Refs. \cite{by,tan,fur5}, the expression for the total persistent currents is given by $\mathcal{I}=\sum_{n,\,l}\mathcal{I}_{n,\,l}$, where $\mathcal{I}_{n,\,l}=-\frac{\partial\mathcal{E}_{n,\,l,\,k}}{\partial\Phi_{B}}$ is called as the Byers-Yang relation \cite{by,bf2}. Therefore, the persistent current associated with the lowest energy state of the system is given by \cite{bb}
\begin{eqnarray}
\mathcal{I}_{1,\,l}=-\frac{\partial\mathcal{E}_{1,\,l,\,k}}{\partial\Phi_{B}}=\mp\frac{q}{4\pi}\,\frac{\varsigma}{\left|\varsigma\right|}\,\frac{m\left(4\left|\varsigma\right|+7\right)}{\sqrt{\left(2\left|\varsigma\right|+3\right)\left(\left|\varsigma\right|+2\right)+\frac{k^{2}}{m^{2}}}}.
\label{3.5}
\end{eqnarray}

By observing the persistent current associated with the lowest energy state (\ref{3.5}), we can see the influence of torsion on the persistent current. By taking $\chi=0$, then, we obtain the persistent current associated with the lowest energy state in the Minkowski spacetime. Therefore, the presence of torsion in the spacetime means that the effects of torsion change the pattern of oscillations of the persistent currents \cite{bf2}.

\section{Conclusions}

We have investigated the effects of torsion of a relativistic position-dependent mass system by showing that analytical solutions to the Klein-Gordon equation can be achieved. We have also seen, in the search for polynomial solutions to the radial wave function, that there exists a restriction on the values of the parameter associated with the linear scalar potential, where the possible values of this parameter are determined by topology of the spacetime (torsion effects) and the quantum numbers of the system. 

Furthermore, we have investigated the effects of torsion when the relativistic position-dependent mass system is subject to the Coulomb potential. In contrast to the first case analyse, we have seen that the presence of the Coulomb potential modifies the relativistic energy levels of the position-dependent mass system. By searching for polynomial solutions to the radial wave function, we have also observed a restriction on the values of the parameter associated with the linear scalar potential, where its allowed values are established by the quantum numbers of the system and the parameters that characterize the Coulomb potential and the torsion in the spacetime.

Finally, we have considered the presence of the Aharonov-Bohm quantum flux \cite{ab} and shown that there exists the influence of both torsion and the Aharonov-Bohm geometric quantum phase on the relativistic energy levels. Again, by searching for polynomial solutions to the radial wave function, we have shown that the possible values of the parameter associated with the linear scalar potential depend on the Aharonov-Bohm geometric quantum phase, the parameter associated with the torsion of the spacetime and the quantum numbers of the system. In particular, we have shown this dependence on the topology of the spacetime, the quantum numbers and the Aharonov-Bohm geometric quantum phase on the allowed energies of the lowest energy state of the system. From this dependence of the energy levels on the geometric quantum phase, we have calculated the persistent current associated with the lowest energy state of the system.

\acknowledgments{The authors would like to thank the Brazilian agencies CNPq and CAPES for financial support.}

\end{document}